\documentclass[twocolumn,english,prb,showpacs]{revtex4}
\usepackage[T1]{fontenc}
\usepackage[latin9]{inputenc}
\usepackage{amssymb}
\usepackage{graphicx}
\usepackage{color}

\makeatletter

\providecommand{\tabularnewline}{\\}

\makeatother

\usepackage{babel}
\begin{document}

\title{Finite-Size Corrections for Ground States of Edwards-Anderson Spin
Glasses}

\author{Stefan Boettcher and Stefan Falkner}

\affiliation{Physics Department, Emory University, Atlanta, Georgia 30322, USA}
\begin{abstract}
Extensive computations of ground state energies of the
Edwards-Anderson spin glass on bond-diluted, hypercubic lattices are
conducted in dimensions $d=3,\ldots,7$. Results are presented for
bond-densities exactly at the percolation threshold, $p=p_{c}$, and
deep within the glassy regime, $p>p_{c}$, where finding ground-states
is one of the hardest combinatorial optimization problems. Finite-size
corrections of the form $1/N^{\omega}$ are shown to be consistent
throughout with the prediction $\omega=1-y/d$, where $y$ refers to the
{}``stiffness'' exponent that controls the formation of domain wall
excitations at low temperatures. At $p=p_{c}$, an extrapolation for
$d\to\infty$ appears to match our mean-field results for these
corrections. In the glassy phase, however, $\omega$ does not approach
its anticipated mean-field value of $2/3$, obtained from simulations
of the Sherrington-Kirkpatrick spin glass on an $N$-clique
graph. Instead, the value of $\omega$ reached at the upper critical
dimension matches another type of mean-field spin glass models, namely
those on sparse random networks of regular degree called Bethe
lattices.
\end{abstract}


\maketitle
\section{Introduction}
The relevance of mean-field predictions based on the
Sherrington-Kirkpatrick model (SK)\citep{Sherrington75} for the
finite-dimensional Ising spin-glass introduced by Edwards and Anderson
(EA)\citep{Edwards75} has been an issue of extensive
discussions.\citep{Parisi79,parisi:80a,bray:86,Fisher86,Franz94,Marinari98a,dedominicis:98,Young08}
Qualitative predictions of mean-field calculations often are taken for
granted in non-disordered systems. Yet, many questions have been
raised\citep{bray:86,Fisher86,Marinari98a,krzakala:01,katzgraber:03f,Young04,joerg:08}
about those predictions for the phase diagram of EA obtained from SK,
as solved with replica symmetry breaking (RSB) by
Parisi.\citep{Parisi79,parisi:80a} Whatever the true nature of the
broken symmetry in a real Ising glass may turn out to be, one would
expect tantalizing insights into the properties of energy landscapes
of disordered systems generally\citep{Wales03} by understanding how
any mean-field RSB-solution morphs into its finite-dimensional
counterpart across the upper critical dimension, here, $d_{u}=6$.
Unfortunately, a direct study of this connection with an expansion in
$\epsilon=d_{u}-d$ is beset with technical
difficulties\citep{dedominicis:98} and results remain hard to interpret.  Numerical
approaches have renewed interest in this connection, based on
simulations of 1d long-range
models\citep{katzgraber:03f,Katzgraber03,Katzgraber05b,Leuzzi08} or
of bond-diluted hypercubic
lattices,\citep{Boettcher04b,Boettcher04c,Boettcher05d,Boettcher07a}
which we adapt for our study here. Prior to the development of this
method, scaling studies on ground states beyond $d=4$
($L\leq7$)\citep{Hartmann99b} were unheard of.

In this Letter we investigate finite-size corrections (FSC) in EA on
dilute lattices at some bond-density $p$ and sizes $N=L^d$ in up to $d=7$
dimensions and probe their connection with similar studies in
mean-field models, such as SK or sparse regular random graphs
frequently called Bethe lattices
(BL).\citep{mezard:01,Mezard03,Boettcher03a,Boettcher10a}  For
lattices at the bond-percolation threshold, the exponent $\omega$
adheres to the predicted scaling relation,\citep{Campbell04}
Eq.~(\ref{eq:omegaD}) below, and
connects smoothly with its mean-field counterpart above $d_l$. Most of
our computational effort is directed at showing that, in the glassy
phase, $\omega$ is equally consistent with the same scaling
relation. Although widely conjectured, a rigorous argument
for that relation at $T=0$ when $T_c>0$ is lacking.  The extrapolation
to its apparent mean-field limit is surprising: it does not match up
(nor cross over\footnote{A cross-over between two correction terms of order
  $1/N^{\omega}$ and $1/N^{2/3}$ at $d_{u}$ is ruled out since
  $\omega>2/3$ for all $d$.})  with corrections found numerically for
either, SK\citep{EOSK,Aspelmeier07,Boettcher10b} or BL with $\pm
J$-bonds.\citep{Boettcher03a} Only FSC for BL with
Gaussian bonds\citep{Boettcher10a} seem to provide a plausible
mean-field limit for $\omega$ in EA. This evidence for a
disconnect in the FSC between EA and SK beyond the upper critical
dimension calls upon a better understanding of their relation.

On a practical level, understanding the FSC is an essential ingredient
to infer correct equilibrium properties in the thermodynamic limit
from simulations of (inevitably) small system size
$N$.\citep{Palassini00}  Similarly, controlling the finite-size
behavior is crucial to prove the existence of averages in random graph
and combinatorial optimization problems.\citep{Bollobas} 

\section{Finite Size Corrections in Spin Glasses}
In this
study we focus on the paradigmatic EA Ising spin glass defined by the
Hamiltonian\citep{Edwards75}
\begin{equation}
H=-\sum_{\left\langle i,j\right\rangle }J_{i,j}\sigma_{i}\sigma_{j},\label{eq:Hamiltonian}
\end{equation}
with spin variables $\sigma_{i}\in\left\{\pm1\right\}$ coupled in a
hyper-cubic lattice of size $N=L^{d}$ via nearest-neighbor bonds
$J_{i,j}$, randomly drawn from some distribution $P\left(J\right)$ of
zero mean and variance $\left\langle J^{2}\right\rangle
=J_{0}^{2}$. The difficulty of computing with any statistical accuracy
the FSC of $H$ at $T=0$ traces mainly to two disorder-specific
complications: (1) Averages over many instances have to be taken, and
(2) finding zero- or low-temperature states for each instance with
certainty requires algorithms of exponential complexity. Thus, even
with good heuristic algorithms,\citep{Dagstuhl04} attainable systems
sizes are typically rather limited such that the form of corrections
often must be guessed.\citep{Palassini00,Boettcher07b} Sometimes,
theoretical arguments, such as those valid at $T=T_c$ for
SK,\citep{parisi:93} can be used to justify the extrapolation of
data. But there are no predictions yet for FSC below $T_c$ at
mean-field level. For EA the conjectured scaling relation at $T=0$
(e.g., see Ref.~[\onlinecite{Bouchaud03}]) is
\begin{equation}
\omega=1-\frac{y}{d}.
\label{eq:omegaD}
\end{equation}
For square lattices at 
$T=0=T_{c}$ this was shown in Ref.~[\onlinecite{Campbell04}].  
Eq.~(\ref{eq:omegaD}) relies on the fundamental exponent\citep{Bray84,bray:86}
$y(=\theta)$ governing the {}``stiffness'' of
domain walls in low-temperature excitations.  It is far more
complicated to verify this relation in the glassy phase at $T=0$ for
systems with $T_{c}>0$, such as EA in $d\geq3$.\citep{Bouchaud03} The
smallness of the stiffness exponent $y$ in $d=3$ obscures the
distinction between bulk- and domain-wall effects in the FSC.  While
$y/d$ is found to increase in higher dimensions,\citep{Boettcher05d}
computational limits on the number of degrees of freedom, and hence on
the length $L$, in higher-$d$ simulations diminishes any such
advantage.

Here, we utilize diluted lattices to attain sufficiently large lengths
$L$, in particular, in larger $d$ to access the optimal scenario to
distinguish scaling behaviors. Using the Extremal Optimization (EO)
heuristic\citep{Boettcher01a,Dagstuhl04} to sample many instances at
$T=0$, we find  evidence in dimensions $d=4,\ldots,7$ that the scaling behavior, controlled by domain-wall excitations, also
applies when $T<T_{c}$.\citep{Bouchaud03} 

\subsection{Cost versus Energy}
In ground-states simulations of finite-degree systems (like EA or BL)
it is convenient to measure the ``cost'' $C_0$, i.e., the
\emph{absolute} sum of unsatisfied bond-weights. The Hamiltonian for
EA in Eq.~(\ref{eq:Hamiltonian}) is a sum over all bonds, where this
cost $C$ contributes positively to the energy and satisfied bonds
provide a sum $S$ that contributes negatively, $E=C-S$. The absolute
weight of all bonds, $B=C+S$ is easily counted for each instance,
amounting to a simple relation between cost and energy,
\begin{equation}
E  =  -B+2C.\label{eq:Eis2C-B}
\end{equation}
Averaging over undiluted ($p=1$) lattices with discrete $\pm
J_{0}$-bonds, the absolute weight of all bonds is fixed, $B=dNJ_{0}$,
and $E$ and $C$ fluctuate identically.  In contrast, for randomly
diluted and/or a continuously distributed bonds, instances fluctuate
normally around the ensemble mean, $B\sim \left\langle B\right\rangle
+O\left(\sqrt{N}\right)$ with $\left\langle B\right\rangle=
pdN\left\langle \left|J\right|\right\rangle$ here.  When only a small
fraction of bonds contribute to $C$, as in these dilute systems, those
trivial fluctuations in $B$ contribute a statistical error to the
measurement of $E$ but \emph{not} of $C$, although both exhibit the same
FSC.\citep{Boettcher10a}  Thus,  we prefer to  extrapolate
for the cost density
\begin{equation}
\left\langle c_{0}\right\rangle_N\equiv\frac{\left\langle C_{0}\right\rangle_N}{N}
\sim \left\langle c_{0}\right\rangle_{\infty} +\frac{1}{2}\Upsilon N^{-\omega},
\label{eq:c0Lscaling}
\end{equation}
and use Eq.~(\ref{eq:Eis2C-B}) with the energy density in the
thermodynamic limit, $\left\langle e_{0}\right\rangle_{\infty}=2\left\langle
c_{0}\right\rangle_{\infty}-\left\langle B\right\rangle/N$, to obtain
\begin{equation}
\left\langle e_{0}\right\rangle\equiv\frac{\left\langle
  E_{0}\right\rangle_N}{N}\sim \left\langle
e_{0}\right\rangle_{\infty} +\Upsilon N^{-\omega}.
\label{eq:e0Lscaling}
\end{equation}
When $C_0$ is intensive, in fact, \emph{only} a measurement of $C$ can
provide the relevant FSC for $E$: Now $\left\langle
C_{0}\right\rangle_N\sim \left\langle C_{0}\right\rangle_{\infty}
+\frac{1}{2}\Upsilon N^{1-\omega}$ leads to Eq.~(\ref{eq:e0Lscaling})
with $\left\langle e_{0}\right\rangle_{\infty}\sim \left(-\left\langle
B\right\rangle+2\left\langle C_{0}\right\rangle_{\infty}\right)/N$. A
direct measurement of $\left\langle e_{0}\right\rangle$ would merely
yield the trivial constant $-\left\langle B\right\rangle/N$ with
$2\left\langle C_{0}\right\rangle_{\infty}/N$ as equally trivial
``correction''.  Note that any such finite-degree systems is very
different from SK, for which $\left\langle B\right\rangle\sim J_0N^2/2+O(N)$
and $\left\langle C_0\right\rangle\sim
N^2/4\left(J_0-2e_{SK}/\sqrt{N}\right)$ so that by
Eq.~(\ref{eq:Eis2C-B}) $e_{SK}\sim\left\langle E\right\rangle/N^{3/2}$
remains as ground-state energy density.

\subsection{Finite-Size Corrections from Domain Wall}
It is well-known that at low temperatures thermal excitations in Ising
spin systems take the form of {}``droplets'',\citep{bray:86,Fisher86}
which at size $s$ require a fluctuation in the energy $\sim s^{y}$,
due to the build-up of interfacial energy. For $y>0$, large droplets
are inhibited, indicating the existence of an ordered state, i.e., a
phase transition must occur at some $T_{c}>0$. For instance, in an
Ising ferromagnet, droplets must be compact to minimize interfacial
energies at $\sim s^{d-1}$, i.e., $y=d-1$, making $d_l=1$ the lower
critical dimension. In EA, such droplets are shaped irregularly to
take advantage of the bond-disorder, bounding
$y<(d-1)/2$,\citep{Fisher86} and in fact, it is
$d_l=2.5$.\citep{Bray84,Franz94,Boettcher05d}

\begin{table}[b!]
\caption{\label{tab:Stiffness-exponents}
Stiffness exponents for Edwards-Anderson spin glasses for dimensions
$d=3,\ldots,7$ obtained numerically from domain-wall excitations of
ground states. The exponent $y_{P}$ refers to lattices at the
bond-percolation threshold $p_{c}$,\citep{Boettcher07a} and $y$
characterizes the $T=0$ fixed point in the glassy phase (since
$T_{c}>0$ for $d\geq3$) for bond densities
$p_{c}<p\leq1$.\citep{Boettcher04c} The last column contains the
measured values, denoted as $\bar\omega$, from the fit in Fig.~\ref{cost_allD}.}
\begin{tabular}{|c||l|l||l|l||c|}
\hline 
$d$ & $y_{P}$[\onlinecite{Boettcher07a}] & $1-y_{P}/d$ &
$y$[\onlinecite{Boettcher04c}] & $1-y/d$ & $\bar\omega$\tabularnewline
\hline 
\hline 
3 & -1.289(6) & 1.429(3) & 0.24(1) & 0.920(4)&0.915(4)\tabularnewline
\hline 
4 & -1.574(6) & 1.393(2) & 0.61(1) & 0.847(3)&0.82(1)\tabularnewline
\hline 
5 & -1.84(2) & 1.368(4) & 0.88(5) & 0.824(10)&0.81(1)\tabularnewline
\hline 
6 & -2.01(4) & 1.335(7) & 1.1(1) & 0.82(2)&0.82(2)\tabularnewline
\hline 
7 & -2.28(6) & 1.33(1) & 1.24(5) & 0.823(7)&0.91(5)\tabularnewline
\hline 
\end{tabular}
\end{table}

For EA at any bond-density $p$, the ground state energies are
extensive on a finite-dimensional lattice, $E_{0}\left(L\right)\sim
L^{d}=N$. With periodic boundary conditions,\citep{Campbell04} the
FSC to this behavior is believed to arise primarily from frustration
imposed by the bond disorder that manifests itself via droplet-like
interfaces spanning the system.  (An example are the {}``defect
lines'' connecting frustrated plaquettes in 2d.) Hence,
\begin{equation}
\left\langle E_{0}\right\rangle _{L}\sim\left\langle
e_{0}\right\rangle _{\infty}L^{d}+\Upsilon L^{y},\qquad(L\to\infty).
\label{eq:E0scaling}
\end{equation}
Rewriting Eq.~(\ref{eq:E0scaling}) in terms of the energy density
$\left\langle e_{0}\right\rangle =\left\langle E_{0}\right\rangle
/L^{d}$ then yields Eq.~(\ref{eq:e0Lscaling}) with the FSC exponent
$\omega$ given by Eq.~(\ref{eq:omegaD}), which is derived easily below
$d_l$ with $T_{c}=0$, i.e., $y\leq0$.

Extending Eq.~(\ref{eq:E0scaling}) naively above $d_l$ for $0\leq
T<T_{c}$, i.e., $y>0$, has not been tested with sufficient data even
in $d=3$.\citep{Pal96a,Hartmann97,Bouchaud03}  Since finding ground
states becomes a hard combinatorial problem requiring heuristics,
attainable system sizes $L$ have been too small to savely distinguish
domain-wall generated from higher-order corrections such as bulk
effects ($\omega=1$), since $y$ is small. E.g., in $d=3$, the lowest
dimension above $d_l$, for which the largest systems sizes ($L\leq12$)
are accessible, we have $y=0.24(1),$ i.e., $y/d\approx0.08$ and
$\omega=1-y/d\approx0.92$.

Using bond-diluted lattices, we have previously determined accurate
values for $y$ from the domain-wall energy in
$d=3,\ldots,7$.\citep{Boettcher04b,Boettcher04c}  For one, dilute
lattices provide large length scales $L$ and higher $d$ for which
ground states can still be obtained in good approximation.  For our
purposes, a higher $d$ also offers a more distinct ratio for
$y/d,$ see Tab.~\ref{tab:Stiffness-exponents}.  Already in $d=4$ it is $y=0.61(1)$, i.e.,
$y/4\approx0.15,$ about \emph{double} the value in $d=3$. 

\begin{figure}
\includegraphics[bb=0bp 400bp 600bp 700bp,angle=270,scale=0.18]{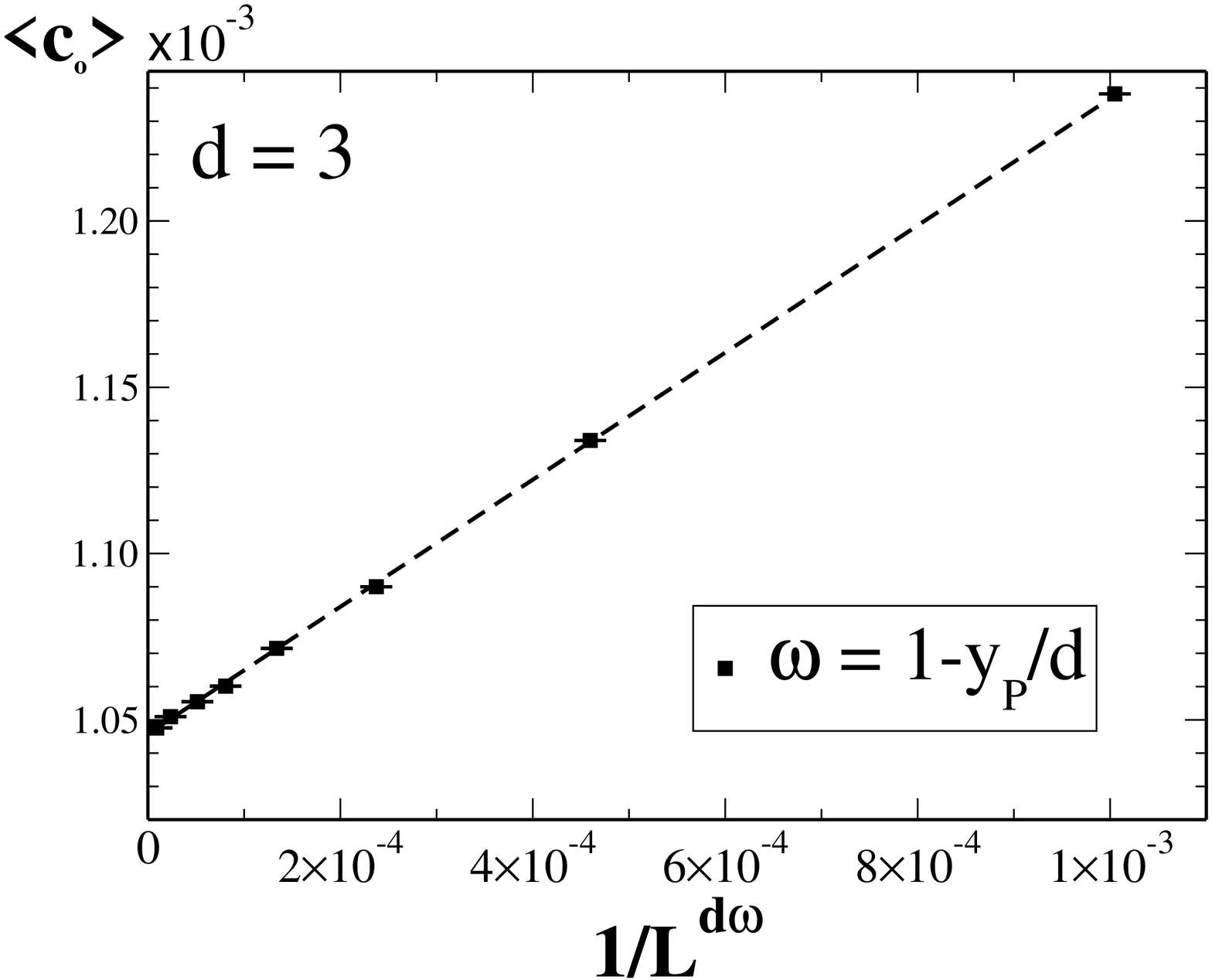}
\includegraphics[bb=0bp 20bp 600bp 400bp,angle=270,scale=0.18]{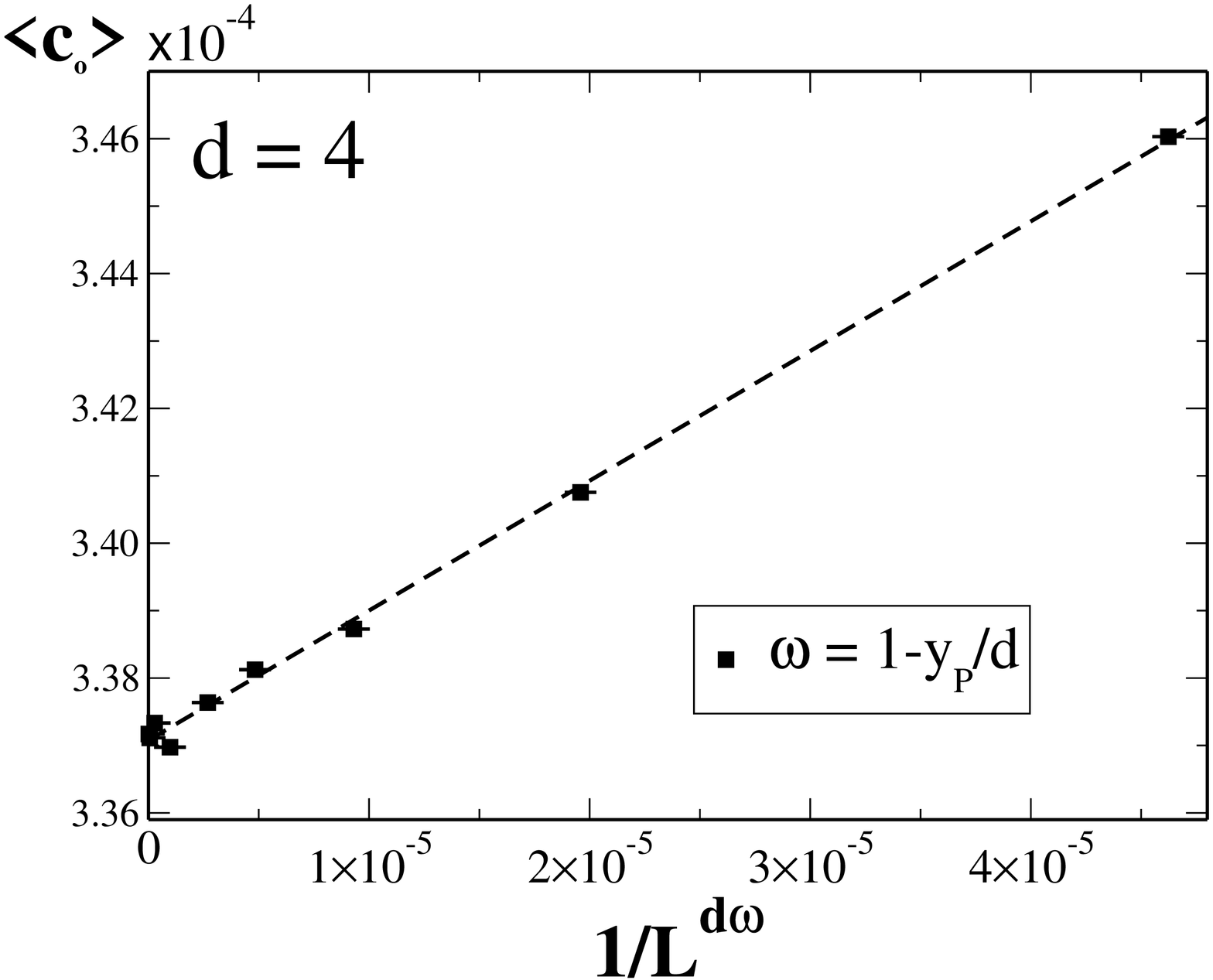}

\caption{\label{cost_pc}
Plot of the approximate ground state cost per spin, $\left\langle
c_{0}\right\rangle $, at bond-percolation $(p=p_{c})$ in dimension
$d=3$ (top) and 4 (bottom) on lattices up to $L\leq100.$ The data is
plotted with respect to the presumed FSC, $1/L^{d\omega}$ with
$\omega=1-y_{P}/d$, using $y_{P}=-1.29(1)$ in $d=3$ and
$y_{P}=-1.574(6)$ in $d=4$, see
Tab.~\ref{tab:Stiffness-exponents}. Data for $L<7$ (way off scale to
be shown) deviates from the asymptotic scaling.}
\end{figure}

\begin{figure}
\includegraphics[bb=0bp 0bp 600bp 740bp,angle=270,scale=0.32]{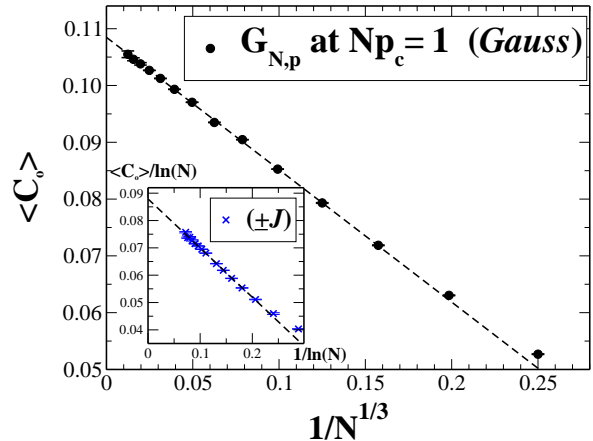}
\caption{\label{costER}
Plot of the average ground state cost $\left\langle
C_{0}\right\rangle $ as a function of $1/N^{1/3},$ obtained with the
reduction method described in Refs.~[\onlinecite{Boettcher04b,Boettcher04c}],
for percolating random graphs $G_{N,p}$ (degree $Np=1)$ of sizes up to
$N=2^{19}$, using a continuous (Gaussian) bond
distribution. Apparently, the total cost itself is intensive and
approached with $\left\langle C_{0}\right\rangle _{N}\sim\left\langle
C_{0}\right\rangle _{\infty}+\frac{1}{2}\Upsilon/N^{1/3}$. Inset:
Corresponding plot for discrete bonds, exhibiting entirely distinct
scaling, $\left\langle C_{0}\right\rangle _{N}\sim\left\langle
C_{0}\right\rangle _{\infty}\ln(N)+\Upsilon/2$.}
\end{figure}

\begin{figure}
\includegraphics[bb=50bp 0bp 612bp 750bp,clip,angle=270,scale=0.32]{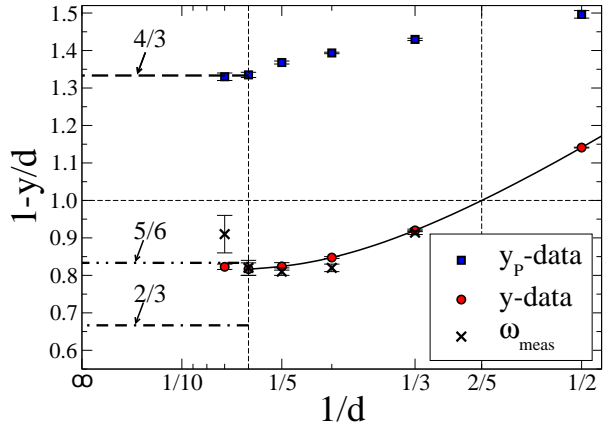}

\caption{\label{thetaoverD}Plot of $1-y/d$ and the measured values, 
$\bar\omega$, in Tab.~\ref{tab:Stiffness-exponents}
as a function of inverse dimension $1/d$. The vertical lines marks
the upper and lower critical dimensions of EA at $d_{u}=6$ (left)
and $d_l\approx5/2$ (right), at which $y=0$. Possible mean-field
predictions to hold above $d_{u}$ are marked by horizontal lines:
$\omega=4/3$ for lattices at $p=p_{c}$, $\omega=5/6$ for EA-lattices
at $p>p_{c}$, and $\omega=2/3$ as predicted from simulations of
SK. The extrapolations in Figs.~\ref{cost_pc} and~\ref{cost_allD}
suggest the validity of Eq.~(\ref{eq:omegaD}) for all $d$, and it
seems difficult to relate the FSC in SK to those of EA. }
\end{figure}

\section{Edwards-Anderson Model at the Percolation Threshold}
As a demonstration of our approach, we first treat the problem on
strongly diluted EA-lattices exactly at $p=p_{c}$, previously studied
in Ref.~[\onlinecite{Boettcher07a}]. The fractal lattice at $p_{c}$ is too
ramified to sustain order and $y=y_{P}<0$ for all
$d$.\citep{Banavar87}  Polynomial-time algorithms have been used to
achieve system sizes up to $N\approx10^{8}$, i.e., $L\leq11$ in $d=7$,
and the results for $y_{P}$ are recounted in
Tab.~\ref{tab:Stiffness-exponents}.  With the available
data, we focus on demonstrating its consistency with Eq.~(\ref{eq:omegaD})
by plotting it on the appropriate scale. 
The results for the extrapolation of the ground state cost for
increasing system size $L$ according to Eq.~(\ref{eq:c0Lscaling}) are
shown in Fig.~\ref{cost_pc} for  $d=3$ and~4. Indeed, the linearity of the plot on the
scale $1/L^{d\omega}$ with $\omega=1-y_{P}/d$, as given in
Tab.~\ref{tab:Stiffness-exponents} for $p=p_{c}$, supports
Eq.~(\ref{eq:omegaD}), as expected for $T_{c}=0$ and
$y_{P}<0$.\citep{Campbell04} However, it is interesting to note that
the data in Tab.~\ref{tab:Stiffness-exponents} suggests $\omega\to4/3$
for $d\to\infty$, as depicted in Fig.~\ref{thetaoverD}. In the
large-$d$ limit, a lattice at $p=p_{c}\sim1/(2d)$ is similar to an
ordinary random graph\citep{Bollobas} $G_{N,p}$ at percolation,
$Np\to1$,\footnote{The lattice does have, in fact, a Poissonian degree
  distribution with an average degree $2dp_{c}\to1$ for
  $d\to\infty$.} this would predict FSC of the form $1/N^{4/3}$ for
the cost density of percolating $G_{N,p}$. As Fig.~\ref{costER} shows,
we do find an FSC with $\omega=4/3$ when using Gaussian bonds, except
for the fact that the cost $\left\langle C_{0}\right\rangle $ itself
in such a limit has become intensive. The energy remains trivially
extensive, $E_{0}\sim-B\gg C_{0}$, as discussed above.

{}Fig.~\ref{costER} also shows the corresponding result using a
discrete ($\pm J_{0}$) bond distribution but the same graph
ensemble. Note that the change in distribution affects an entirely
different scaling. The leading-order difference can be explained by
envisioning $G_{N,p}$ near percolation as a tree possessing
$O\left(\ln N\right)$ weakly interlinked loops, each of length
$O\left(\ln N\right)$, with half of those (1d-like) loops frustrated
in a single bond: The total cost for discrete bonds of \emph{fixed}
weight $\left|J\right|=J_{0}$ is $\left\langle C_{0}\right\rangle
_{N}\sim J_{0}\ln N$, while for continuous bonds this cost is
mitigated by selecting the \emph{weakest} weight, $\left|J\right|\sim
J_0/\ln N$, in the loop and  $\left\langle C_{0}\right\rangle
_{N}$ approaches a constant value.

We conclude for $p=p_{c}$ that our results in finite $d$ connect well
with those for $d\to\infty$, as long as we choose a continuous bond
distribution in the mean-field limit. While non-universality is not
unexpected when $T_{c}=0$,\citep{Hartmann01} we will encounter it also
for FSC when $T_{c}>0$ because FSC do not affect the thermodynamic
limit.

\section{Edwards-Anderson Model in the Glassy State}
Parallel to the above case, we analyze our data for dilute EA-lattices
in the glassy regime, $p>p_{c}$. However, this data had to be obtained
in far more laborious simulations. While starting from lattices of
comparable size, up to $N\sim10^{8}$, the methods from
Ref.~[\onlinecite{Boettcher07a}] merely reduce each instance to a remainder
of about $N_{R}\lesssim10^{3}$ variables at the chosen values of
$p>p_{c}$. (The choice of $p$ is crucial: larger values dramatically
increase the remainder sizes, smaller values create long transients
due to the proximity of $p_{c}$.)  The cost $C_0$ of each remainder
had to be optimized by local search with EO,\citep{Boettcher01a} run
with $O\left(N_{R}^{3}\right)$ updates.\citep{Boettcher04b} Depending
on system size $L$ and dimension $d$, disorder averaging required
$10^{4}-10^{7}$ instances to reduce statistical errors.  In contrast
to our earlier simulations,\citep{Boettcher04b,Boettcher04c} far more
statistical accuracy  for each energy average and new
implementations to handle very large lattices efficiently are required here.\footnote{In
  all, about four months of computations on a 50-node cluster were
  used for this project.}  While there is never a guarantee for
finding actual ground states, we have applied similar safeguards to
hold systematic errors below statistical errors as in previous
investigations.\citep{Boettcher01a,Boettcher03a,Dagstuhl04,Boettcher10a} For
instance, in $d=7$ at $L=11$ we have tested subsets of 100 instances
at run-times of twice the normal setting and did not find a single
deviation.

\begin{figure*}
\includegraphics[bb=275bp 0bp 580bp 720bp,clip,angle=270,scale=0.34]{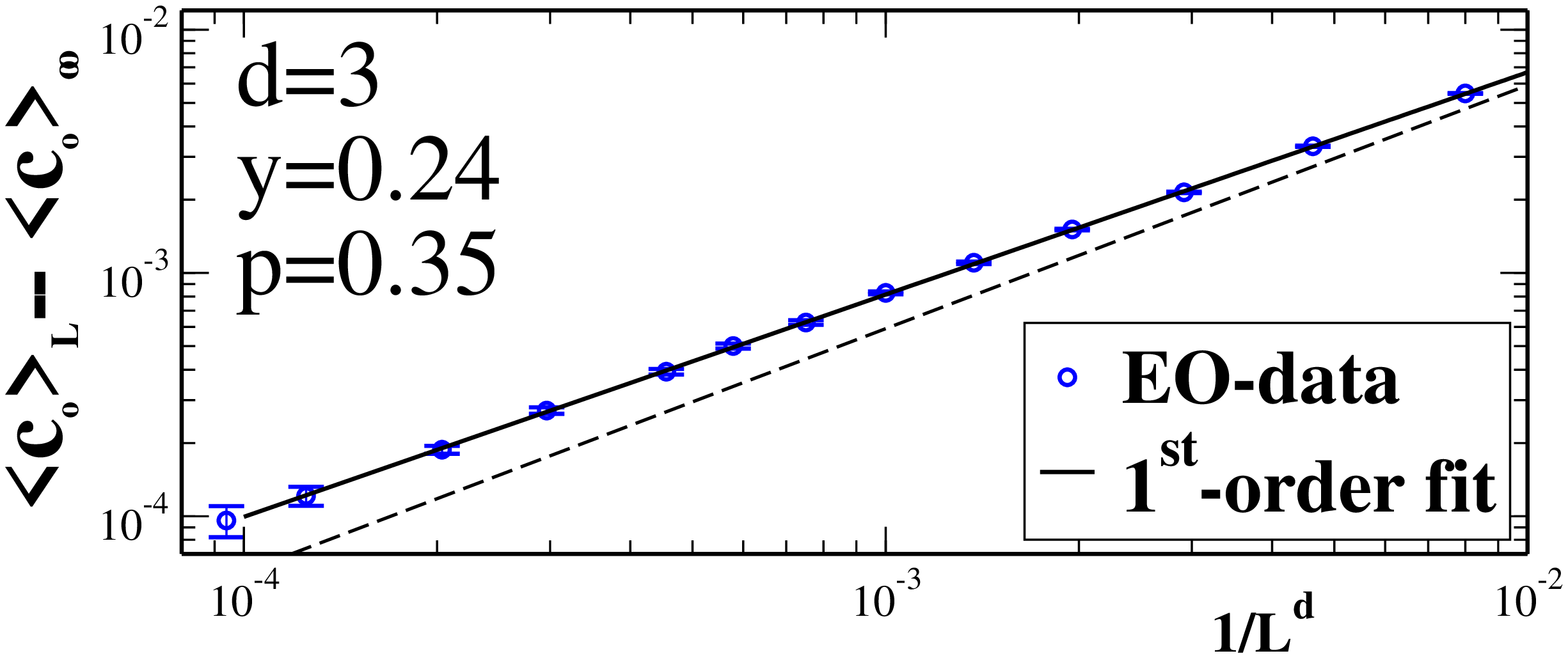}
\includegraphics[bb=275bp 0bp 580bp 720bp,clip,angle=270,scale=0.34]{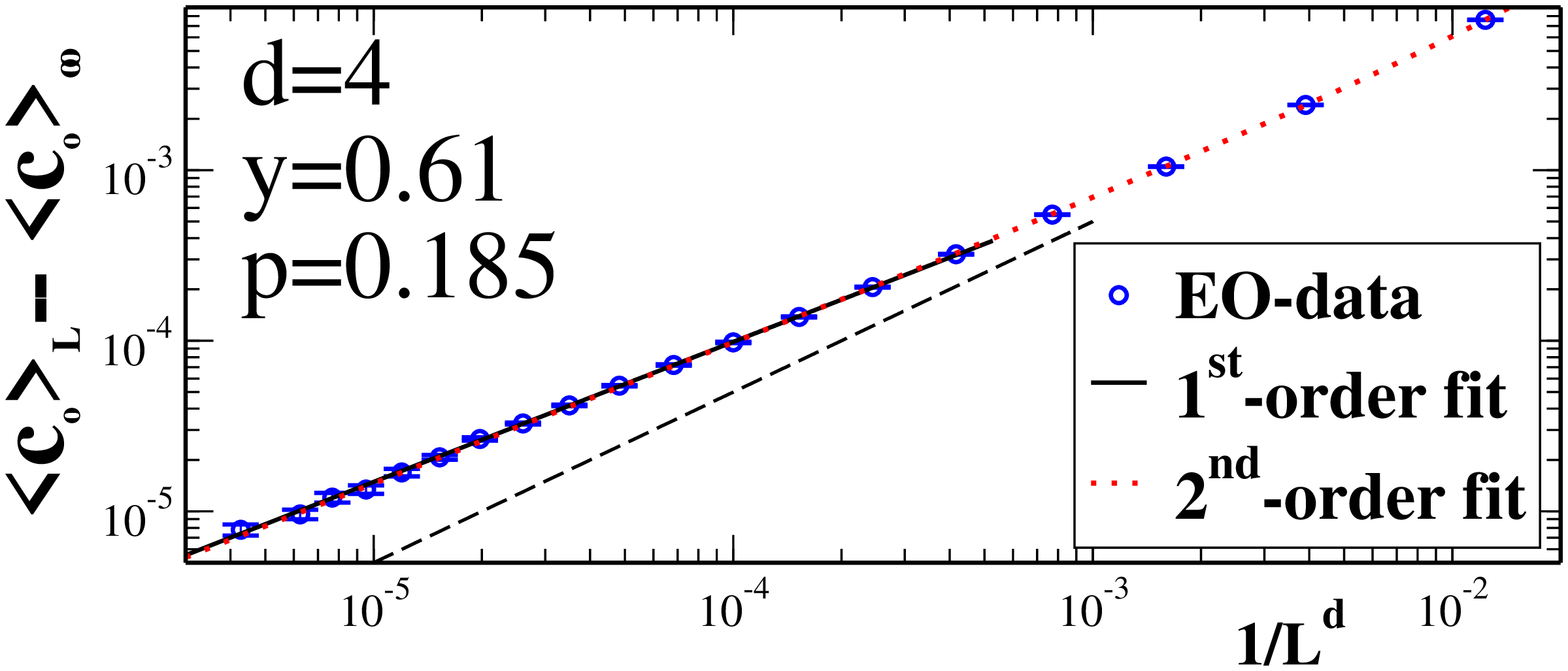}

\includegraphics[bb=275bp 0bp 580bp 720bp,clip,angle=270,scale=0.34]{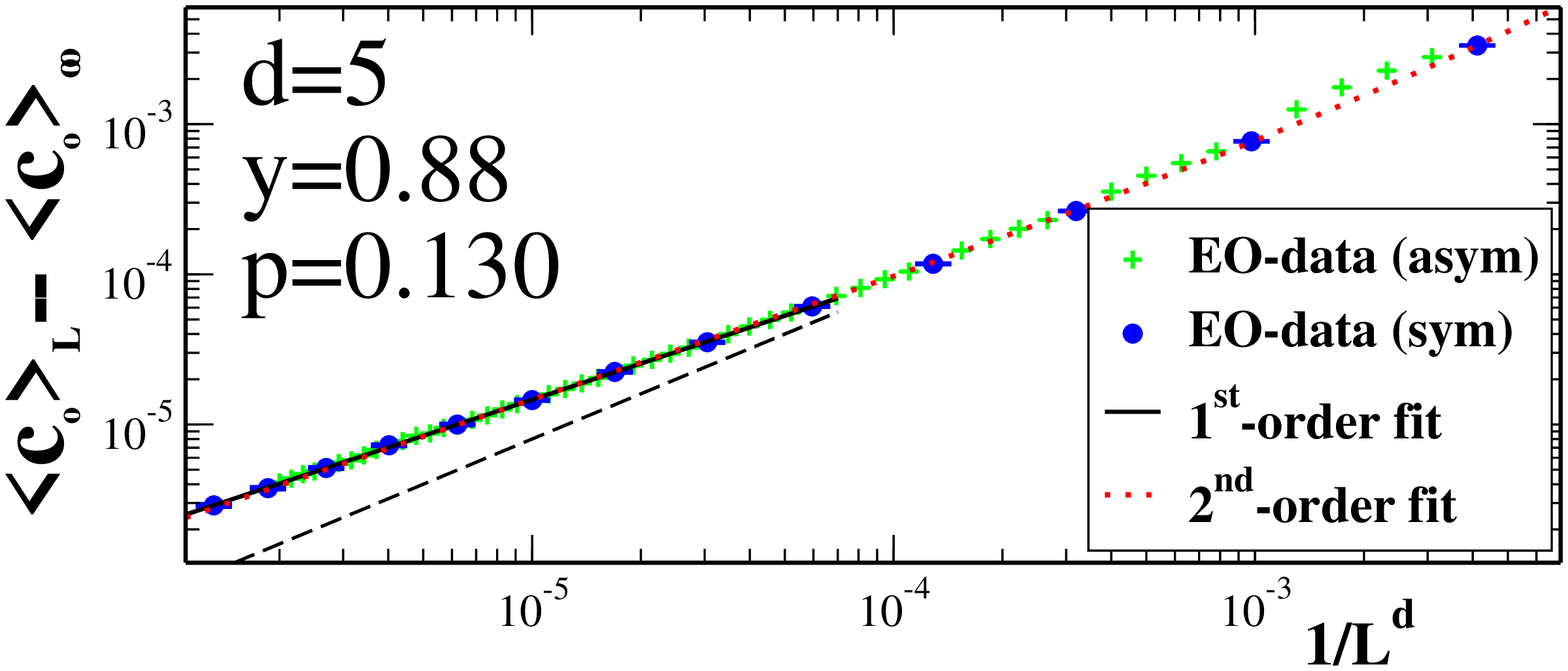}
\includegraphics[bb=275bp 0bp 580bp 720bp,clip,angle=270,scale=0.34]{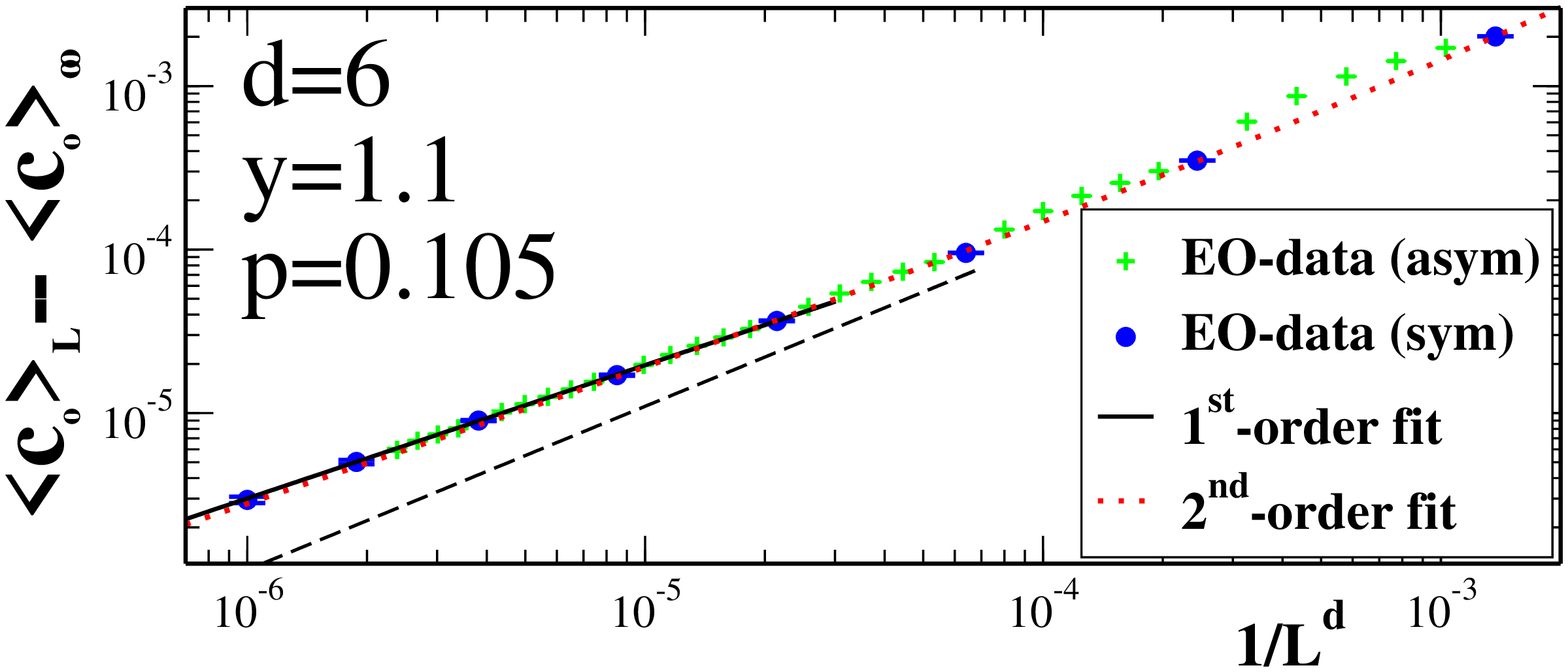}

\includegraphics[bb=275bp 0bp 570bp 720bp,clip,angle=270,scale=0.34]{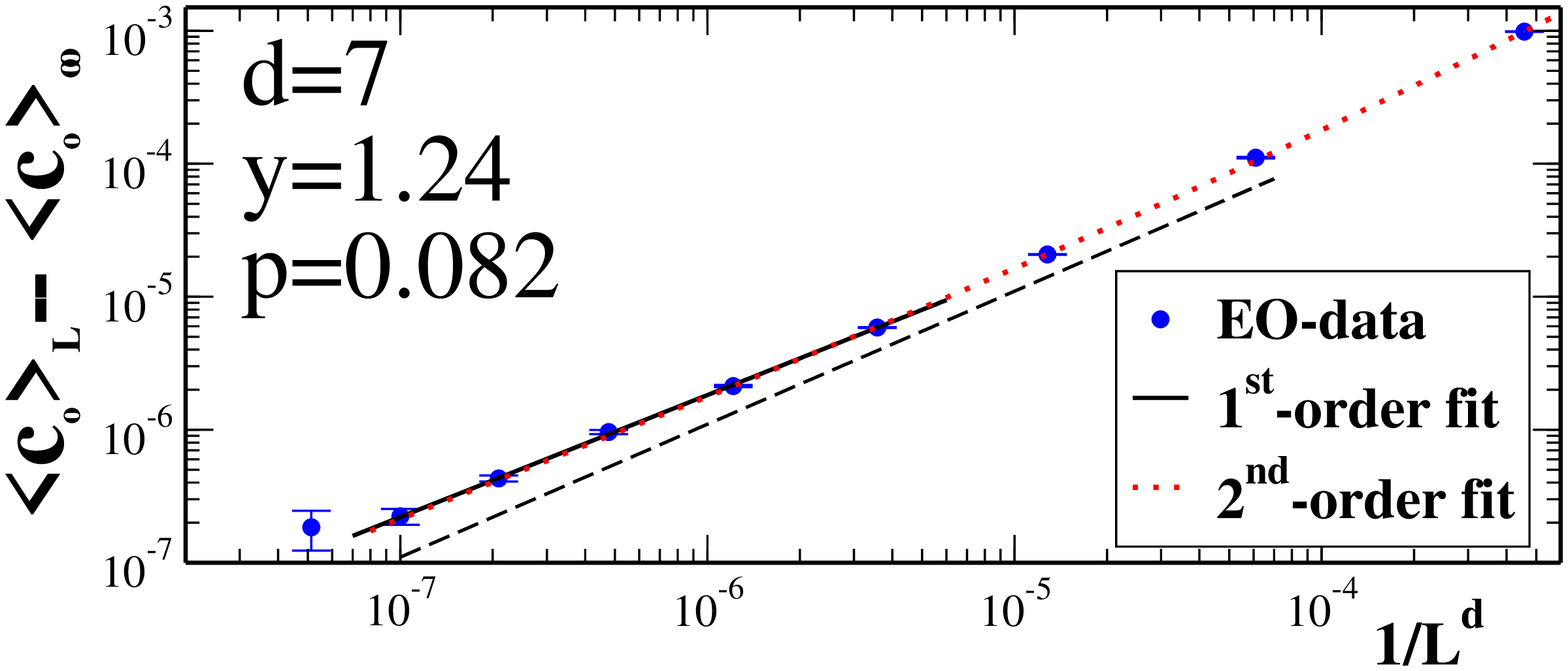}
\caption{\label{cost_allD}Double-logarithmic plots of the approximate
  ground state cost per spin, $\left\langle c_{0}\right\rangle
  _{L}=\left\langle C_{0}\right\rangle _{L}/L^{d}$, reduced by its numerically
  extrapolated limit value $\left\langle c_{0}\right\rangle
  _{\infty}=\lim_{L\to\infty}\left\langle c_{0}\right\rangle
  _{L}$, versus $1/L^d$ for
  dimensions $d=3,\ldots,7$. Subtracting off $\left\langle c_{0}\right\rangle
  _{\infty}$ highlights the FSC-term in Eq.~(\ref{eq:c0Lscaling}). Simulated lattice sizes range from $L=3$
  to $L\leq25$ ($d=3$), $L\leq20$ ($d=4$), $L\leq15$ ($d=5$), and $L\leq10$ ($d=6,7$) at
  respective bond densities $p=0.185$ ($d=4$), $p=0.13$ ($d=5$),
  $p=0.105$ ($d=6$), and $p=0.082$ ($d=7$). The $1^{\rm st}$-order
  fit (solid line) to obtain $\left\langle c_{0}\right\rangle
  _{\infty}$ and $\bar\omega$ using Eq.~(\ref{eq:c0Lscaling}) only
  utilized the fitted data-points. The measured values for $\bar\omega$
  are reported in Tab.~\ref{tab:Stiffness-exponents}. The $2^{\rm nd}$-order
  fit (dotted line), involving \emph{all} data for $d=4,\ldots,7$,  held previously
  fitted parameters fixed but added a further correction term to
  Eq.~(\ref{eq:c0Lscaling}). This correction shows little effect on
  the asymptotic scaling, as solid and dotted lines virtually
  coincide. The dashed line for any ``bulk''
  corrections of order $1/L^d$ is provided for reference only.
Data (\textcolor{green}{+}) in $d=5,6$ obtained from asymmetric
instances (labeled {}``asym'') was exclude from the fits; it is
displayed to demonstrate its consistency, being systematically ``arced'' above
the symmetric data.}
\end{figure*}

To avoid the above-mentioned problem of wide separation between data
points at $L$ and $L+1$ in higher $d$, we have also undertaken to
simulate \emph{asymmetric} instances of size $N=L^{d-m}(L+1)^{m}$ for
$m=1,2,\ldots,d-1$. While we display that data as well, it proved
unsuitable to fit. It merely serves to demonstrate that even such a
minor variation in rather large, random instances resulted in
sufficiently large systematic distortions from the $m=0$ data to be
detectable with our heuristic.

The results of this effort for $d=3,\ldots,7$ are presented in
Fig.~\ref{cost_allD}. In each case, the raw data in the asymptotic
scaling regime has been fit to the form in Eq.~(\ref{eq:c0Lscaling})
to yield  $\bar{\omega}=0.914(5)$ and $\left\langle
c_{0}\right\rangle_{\infty}\approx0.0425063$ with
$\chi^{2}/ndf\approx1.3$ ($d=3$), $\bar{\omega}=0.82(1)$ and $\left\langle
c_{0}\right\rangle _{\infty}\approx0.0063720$ with
$\chi^{2}/ndf\approx0.77$ ($d=4$), $\bar{\omega}=0.81(1)$ and
$\left\langle c_{0}\right\rangle _{\infty}\approx0.0022543$ with
$\chi^{2}/ndf\approx1.8$ ($d=5$), $\bar{\omega}=0.82(2)$ and
$\left\langle c_{0}\right\rangle _{\infty}\approx0.0014851$ with
$\chi^{2}/ndf\approx0.8$ ($d=6$), $\bar{\omega}=0.91(5)$ and
$\left\langle c_{0}\right\rangle _{\infty}\approx0.00061055$
($d=7$). Each fitted value of $\bar\omega$ is close to $1-y/d$, see
Tab.~\ref{tab:Stiffness-exponents}  and Fig.~\ref{thetaoverD}, except for $d=7$. For the largest
$d$, the value of $\bar\omega$ trends higher, although it is difficult
to judge whether asymptotic behavior has been reached.  Fixing
$\omega=1-y/d$ and using a $1/L^{d\omega}$-scale as in Fig.~\ref{cost_pc}, a two-parameter linear fit (not shown) yields
virtually identical values: $\left\langle c_{0}\right\rangle
_{\infty}\approx0.0063735$ ($d=4$), $\left\langle c_{0}\right\rangle
_{\infty}\approx0.0022546$ ($d=5$), $\left\langle c_{0}\right\rangle
_{\infty}\approx0.0014852$ ($d=6$), and $\left\langle
c_{0}\right\rangle _{\infty}\approx0.00061047$ ($d=7$).  In
Fig.~\ref{cost_allD} we have subtracted the fitetd value of
$\left\langle c_{0}\right\rangle _{\infty}$ from each data point to be
able to plot the remaining scaling form on a double-logarithmic plot,
for better visibility.

To assess the effect of higher order corrections on our fits, we fixed
the parameters of the $1^{\rm st}$-order fit\footnote{It is
  impossible to fit the data with two equivalent terms,
  each with its own fitting exponent, in the hope these would converge spontaneously 
  to the first and second order correction.} and added another
scaling correction term to Eq.~(\ref{eq:c0Lscaling}) for a $2^{\rm
  nd}$-order fit, now including all points down to the smallest size,
$L=3$.  In each case, these corrections scaled with an exponent of
$\approx 1.3-1.6$, about \emph{twice} that of the leading
correction. Shown as dotted lines in Fig.~\ref{cost_allD}, these
higher-order corrections have virtually no impact on the asymptotic
behavior. We have tested that the existence of such a rapidly
vanishing correction does not affect our results by more than 2\% in
$\bar\omega$.

\section{Conclusions}
We conclude, first and foremost, that the ground-state FSC in EA are
consistent with Eq.~(\ref{eq:omegaD}), for all $d\leq d_u=6$, whether
$T_c=0$ ($p=p_c$) or $T_c>0$ ($p>p_c$). The connection with mean-field
predictions is more complicated.  The FSC exponent $\omega$ for EA
does not approach its equivalent of $\omega_{SK}\approx2/3$ in
SK\citep{EOSK,Aspelmeier07,Boettcher10b,Boettcher03a} at $d_u$, by
far. As Fig.~\ref{thetaoverD} shows, its mean-field limit seems more
aligned with the result of $\omega\approx0.8$ found for sparse-degree
BL having a continuous bond distribution.\citep{Boettcher10a} There
is no obvious reason why those BL would provide a more appropriate mean-field 
limit for EA, except for their sparse degree.  In the computationally simpler 
case of EA at $p=p_c$, a similar approach for FSC to a sparse mean-field limit 
is even more apparent. 

Beyond FSC, ground-state energy fluctuations provide another example
where its exponent $\rho$ in EA does not reach the corresponding SK
value for large $d$. It has been proven\citep{Wehr90} that those
fluctuations are normal (i.e., $\rho=1/2$) for all $d$ in EA, whereas
in SK it has a value of $\rho=5/6$.\citep{Parisi08} (In fact, it has
been argued\citep{Aspelmeier03} that $\rho=1-y/d$ for $d>d_u$, for which
Eq.~(\ref{eq:omegaD}) implies $\omega=\rho=5/6$ at $d_u$, consistent with our data in
Tab.~\ref{tab:Stiffness-exponents} and Fig.~\ref{thetaoverD}.) Again,
BL with Gaussian bonds appear to have the right mean-field limit for
EA, exhibiting $\rho=1/2$.\citep{Boettcher10b}

To sort out these behaviors we first note that both, $\omega$ and
$\rho$, merely describe properties of the asymptotic approach to the thermodynamic limit
and, hence, may exhibit non-universal (i.e., distribution-dependent)
behavior, even though they can be related to other, universal exponents such
as $y$. Non-universal, distribution-dependent behavior is far more prevalent for sparse mean-field spin
glasses than for SK.\citep{Boettcher10a,Boettcher10b} For such properties, it ought not
surprise that these sparse systems provide a better high-dimensional
limit for EA than SK. After all, in studying EA we first take the
limit $N=L^d\to\infty$ at fixed $d$ (for a sparse degree of $2d$)
before we consider $d\to\infty$, whereas SK corresponds to the
correlated limit, $N\to\infty$, $d\to\infty$ with degree $2d\sim N$
held fixed. In disordered systems, the order in which those limits are
taken might well matter.  However, the quest to understand the extend or possible
break-down of RSB, which is known to describe the glassy state in
sparse mean-field models\citep{mezard:01,Mezard03} as well as SK,
remains undiminished.

This work has been supported by grant DMR-0812204 from the National
Science Foundation.

\bibliographystyle{apsrev}
\bibliography{/Users/stb/Boettcher}

\begin{thebibliography}{46}
\expandafter\ifx\csname natexlab\endcsname\relax\def\natexlab#1{#1}\fi
\expandafter\ifx\csname bibnamefont\endcsname\relax
  \def\bibnamefont#1{#1}\fi
\expandafter\ifx\csname bibfnamefont\endcsname\relax
  \def\bibfnamefont#1{#1}\fi
\expandafter\ifx\csname citenamefont\endcsname\relax
  \def\citenamefont#1{#1}\fi
\expandafter\ifx\csname url\endcsname\relax
  \def\url#1{\texttt{#1}}\fi
\expandafter\ifx\csname urlprefix\endcsname\relax\def\urlprefix{URL }\fi
\providecommand{\bibinfo}[2]{#2}
\providecommand{\eprint}[2][]{\url{#2}}

\bibitem[{\citenamefont{Sherrington and Kirkpatrick}(1975)}]{Sherrington75}
\bibinfo{author}{\bibfnamefont{D.}~\bibnamefont{Sherrington}} \bibnamefont{and}
  \bibinfo{author}{\bibfnamefont{S.}~\bibnamefont{Kirkpatrick}},
  \bibinfo{journal}{Phys. Rev. Lett.} \textbf{\bibinfo{volume}{35}},
  \bibinfo{pages}{1792} (\bibinfo{year}{1975}).

\bibitem[{\citenamefont{Edwards and Anderson}(1975)}]{Edwards75}
\bibinfo{author}{\bibfnamefont{S.~F.} \bibnamefont{Edwards}} \bibnamefont{and}
  \bibinfo{author}{\bibfnamefont{P.~W.} \bibnamefont{Anderson}},
  \bibinfo{journal}{J. Phys. F} \textbf{\bibinfo{volume}{5}},
  \bibinfo{pages}{965} (\bibinfo{year}{1975}).

\bibitem[{\citenamefont{Parisi}(1979)}]{Parisi79}
\bibinfo{author}{\bibfnamefont{G.}~\bibnamefont{Parisi}},
  \bibinfo{journal}{Phys. Rev. Lett.} \textbf{\bibinfo{volume}{43}},
  \bibinfo{pages}{1754} (\bibinfo{year}{1979}).

\bibitem[{\citenamefont{Parisi}(1980)}]{parisi:80a}
\bibinfo{author}{\bibfnamefont{G.}~\bibnamefont{Parisi}}, \bibinfo{journal}{J.
  Phys. A} \textbf{\bibinfo{volume}{13}}, \bibinfo{pages}{L115}
  (\bibinfo{year}{1980}).

\bibitem[{\citenamefont{Bray and Moore}(1986)}]{bray:86}
\bibinfo{author}{\bibfnamefont{A.~J.} \bibnamefont{Bray}} \bibnamefont{and}
  \bibinfo{author}{\bibfnamefont{M.~A.} \bibnamefont{Moore}}, in
  \emph{\bibinfo{booktitle}{Heidelberg Colloquium on Glassy Dynamics and
  Optimization}}, edited by
  \bibinfo{editor}{\bibfnamefont{L.}~\bibnamefont{Van~Hemmen}}
  \bibnamefont{and}
  \bibinfo{editor}{\bibfnamefont{I.}~\bibnamefont{Morgenstern}}
  (\bibinfo{publisher}{Springer}, \bibinfo{address}{New York},
  \bibinfo{year}{1986}), p. \bibinfo{pages}{121}.

\bibitem[{\citenamefont{Fisher and Huse}(1986)}]{Fisher86}
\bibinfo{author}{\bibfnamefont{D.~S.} \bibnamefont{Fisher}} \bibnamefont{and}
  \bibinfo{author}{\bibfnamefont{D.~A.} \bibnamefont{Huse}},
  \bibinfo{journal}{Phys. Rev. Lett.} \textbf{\bibinfo{volume}{56}},
  \bibinfo{pages}{1601} (\bibinfo{year}{1986}).

\bibitem[{\citenamefont{Franz et~al.}(1994)\citenamefont{Franz, Parisi, and
  Virasoro}}]{Franz94}
\bibinfo{author}{\bibfnamefont{S.}~\bibnamefont{Franz}},
  \bibinfo{author}{\bibfnamefont{G.}~\bibnamefont{Parisi}}, \bibnamefont{and}
  \bibinfo{author}{\bibfnamefont{M.~A.} \bibnamefont{Virasoro}},
  \bibinfo{journal}{J. Phys. I (France)} \textbf{\bibinfo{volume}{4}},
  \bibinfo{pages}{1657} (\bibinfo{year}{1994}).

\bibitem[{\citenamefont{Marinari et~al.}(1998)\citenamefont{Marinari, Parisi,
  and Ruiz-Lorenzo}}]{Marinari98a}
\bibinfo{author}{\bibfnamefont{E.}~\bibnamefont{Marinari}},
  \bibinfo{author}{\bibfnamefont{G.}~\bibnamefont{Parisi}}, \bibnamefont{and}
  \bibinfo{author}{\bibfnamefont{J.~J.} \bibnamefont{Ruiz-Lorenzo}},
  \bibinfo{journal}{Phys. Rev. B} \textbf{\bibinfo{volume}{58}},
  \bibinfo{pages}{14852} (\bibinfo{year}{1998}).

\bibitem[{\citenamefont{de~Dominicis et~al.}(1998)\citenamefont{de~Dominicis,
  Kondor, and Temes{\'a}ri}}]{dedominicis:98}
\bibinfo{author}{\bibfnamefont{C.}~\bibnamefont{de~Dominicis}},
  \bibinfo{author}{\bibfnamefont{I.}~\bibnamefont{Kondor}}, \bibnamefont{and}
  \bibinfo{author}{\bibfnamefont{T.}~\bibnamefont{Temes{\'a}ri}}, in
  \emph{\bibinfo{booktitle}{Spin Glasses and Random Fields}}, edited by
  \bibinfo{editor}{\bibfnamefont{A.}~\bibnamefont{Young}}
  (\bibinfo{publisher}{World Scientific}, \bibinfo{address}{Singapore},
  \bibinfo{year}{1998}).

\bibitem[{\citenamefont{Young}(2008)}]{Young08}
\bibinfo{author}{\bibfnamefont{A.~P.} \bibnamefont{Young}},
  \bibinfo{journal}{J. Phys. A: Math. Theor.}
  \textbf{\bibinfo{volume}{41}}, \bibinfo{pages}{324016}
  (\bibinfo{year}{2008}).

\bibitem[{\citenamefont{Krzakala et~al.}(2001)\citenamefont{Krzakala, Houdayer,
  Marinari, Martin, and Parisi}}]{krzakala:01}
\bibinfo{author}{\bibfnamefont{F.}~\bibnamefont{Krzakala}},
  \bibinfo{author}{\bibfnamefont{J.}~\bibnamefont{Houdayer}},
  \bibinfo{author}{\bibfnamefont{E.}~\bibnamefont{Marinari}},
  \bibinfo{author}{\bibfnamefont{O.~C.} \bibnamefont{Martin}},
  \bibnamefont{and} \bibinfo{author}{\bibfnamefont{G.}~\bibnamefont{Parisi}},
  \bibinfo{journal}{Phys. Rev. Lett.} \textbf{\bibinfo{volume}{87}},
  \bibinfo{pages}{197204} (\bibinfo{year}{2001}).

\bibitem[{\citenamefont{Katzgraber and
  Young}(2003{\natexlab{a}})}]{katzgraber:03f}
\bibinfo{author}{\bibfnamefont{H.~G.} \bibnamefont{Katzgraber}}
  \bibnamefont{and} \bibinfo{author}{\bibfnamefont{A.~P.} \bibnamefont{Young}},
  \bibinfo{journal}{Phys. Rev. B} \textbf{\bibinfo{volume}{68}},
  \bibinfo{pages}{224408} (\bibinfo{year}{2003}{\natexlab{a}}).

\bibitem[{\citenamefont{Young and Katzgraber}(2004)}]{Young04}
\bibinfo{author}{\bibfnamefont{A.~P.} \bibnamefont{Young}} \bibnamefont{and}
  \bibinfo{author}{\bibfnamefont{H.~G.} \bibnamefont{Katzgraber}},
  \bibinfo{journal}{Phys. Rev. Lett.} \textbf{\bibinfo{volume}{93}},
  \bibinfo{pages}{207203} (\bibinfo{year}{2004}).

\bibitem[{\citenamefont{J\"org et~al.}(2008)\citenamefont{J\"org, Katzgraber,
  and Krza\ifmmode~\mbox{\c{}}\else \c{}\fi{}ka\l{}a}}]{joerg:08}
\bibinfo{author}{\bibfnamefont{T.}~\bibnamefont{J\"org}},
  \bibinfo{author}{\bibfnamefont{H.~G.} \bibnamefont{Katzgraber}},
  \bibnamefont{and}
  \bibinfo{author}{\bibfnamefont{F.}~\bibnamefont{Krza\ifmmode~\mbox{\c{}}\else
  \c{}\fi{}ka\l{}a}}, \bibinfo{journal}{Phys. Rev. Lett.}
  \textbf{\bibinfo{volume}{100}}, \bibinfo{pages}{197202}
  (\bibinfo{year}{2008}).

\bibitem[{\citenamefont{Wales}(2003)}]{Wales03}
\bibinfo{author}{\bibfnamefont{D.~J.} \bibnamefont{Wales}},
  \emph{\bibinfo{title}{Energy landscapes}} (\bibinfo{publisher}{Cambridge
  University Press, Cambridge}, \bibinfo{year}{2003}).

\bibitem[{\citenamefont{Katzgraber and
  Young}(2003{\natexlab{b}})}]{Katzgraber03}
\bibinfo{author}{\bibfnamefont{H.~G.} \bibnamefont{Katzgraber}}
  \bibnamefont{and} \bibinfo{author}{\bibfnamefont{A.~P.} \bibnamefont{Young}},
  \bibinfo{journal}{Phys. Rev. B} \textbf{\bibinfo{volume}{67}},
  \bibinfo{pages}{134410} (\bibinfo{year}{2003}{\natexlab{b}}).

\bibitem[{\citenamefont{Katzgraber and Young}(2005)}]{Katzgraber05b}
\bibinfo{author}{\bibfnamefont{H.~G.} \bibnamefont{Katzgraber}}
  \bibnamefont{and} \bibinfo{author}{\bibfnamefont{A.~P.} \bibnamefont{Young}},
  \bibinfo{journal}{Phys. Rev. B} \textbf{\bibinfo{volume}{72}},
  \bibinfo{pages}{184416} (\bibinfo{year}{2005}).

\bibitem[{\citenamefont{Leuzzi et~al.}(2008)\citenamefont{Leuzzi, Parisi,
  Ricci-Tersenghi, and Ruiz-Lorenzo}}]{Leuzzi08}
\bibinfo{author}{\bibfnamefont{L.}~\bibnamefont{Leuzzi}},
  \bibinfo{author}{\bibfnamefont{G.}~\bibnamefont{Parisi}},
  \bibinfo{author}{\bibfnamefont{F.}~\bibnamefont{Ricci-Tersenghi}},
  \bibnamefont{and} \bibinfo{author}{\bibfnamefont{J.~J.}
  \bibnamefont{Ruiz-Lorenzo}}, \bibinfo{journal}{Phys. Rev. Lett.}
  \textbf{\bibinfo{volume}{101}}, \bibinfo{eid}{107203}
  (pages~\bibinfo{numpages}{4}) (\bibinfo{year}{2008}).

\bibitem[{\citenamefont{Boettcher}(2004{\natexlab{a}})}]{Boettcher04b}
\bibinfo{author}{\bibfnamefont{S.}~\bibnamefont{Boettcher}},
  \bibinfo{journal}{Euro. Phys. J. B} \textbf{\bibinfo{volume}{38}},
  \bibinfo{pages}{83} (\bibinfo{year}{2004}{\natexlab{a}}).

\bibitem[{\citenamefont{Boettcher}(2004{\natexlab{b}})}]{Boettcher04c}
\bibinfo{author}{\bibfnamefont{S.}~\bibnamefont{Boettcher}},
  \bibinfo{journal}{Europhys. Lett.} \textbf{\bibinfo{volume}{67}},
  \bibinfo{pages}{453} (\bibinfo{year}{2004}{\natexlab{b}}).

\bibitem[{\citenamefont{Boettcher}(2005{\natexlab{a}})}]{Boettcher05d}
\bibinfo{author}{\bibfnamefont{S.}~\bibnamefont{Boettcher}},
  \bibinfo{journal}{Phys. Rev. Lett.} \textbf{\bibinfo{volume}{95}},
  \bibinfo{pages}{197205} (\bibinfo{year}{2005}{\natexlab{a}}).

\bibitem[{\citenamefont{Boettcher and Marchetti}(2008)}]{Boettcher07a}
\bibinfo{author}{\bibfnamefont{S.}~\bibnamefont{Boettcher}} \bibnamefont{and}
  \bibinfo{author}{\bibfnamefont{E.}~\bibnamefont{Marchetti}},
  \bibinfo{journal}{Phys. Rev. B} \textbf{\bibinfo{volume}{77}},
  \bibinfo{pages}{100405(R)} (\bibinfo{year}{2008}).

\bibitem[{\citenamefont{Hartmann}(1999)}]{Hartmann99b}
\bibinfo{author}{\bibfnamefont{A.~K.} \bibnamefont{Hartmann}},
  \bibinfo{journal}{Phys. Rev. E} \textbf{\bibinfo{volume}{60}},
  \bibinfo{pages}{5135} (\bibinfo{year}{1999}).

\bibitem[{\citenamefont{{M{\'e}zard} and {Parisi}}(2001)}]{mezard:01}
\bibinfo{author}{\bibfnamefont{M.}~\bibnamefont{{M{\'e}zard}}}
  \bibnamefont{and} \bibinfo{author}{\bibfnamefont{G.}~\bibnamefont{{Parisi}}},
  \bibinfo{journal}{Eur. Phys. J. B} \textbf{\bibinfo{volume}{20}},
  \bibinfo{pages}{217} (\bibinfo{year}{2001}).

\bibitem[{\citenamefont{Mezard and Parisi}(2003)}]{Mezard03}
\bibinfo{author}{\bibfnamefont{M.}~\bibnamefont{Mezard}} \bibnamefont{and}
  \bibinfo{author}{\bibfnamefont{G.}~\bibnamefont{Parisi}},
  \bibinfo{journal}{J. Stat. Phys.} \textbf{\bibinfo{volume}{111}},
  \bibinfo{pages}{1} (\bibinfo{year}{2003}).

\bibitem[{\citenamefont{Boettcher}(2003)}]{Boettcher03a}
\bibinfo{author}{\bibfnamefont{S.}~\bibnamefont{Boettcher}},
  \bibinfo{journal}{Euro. Phys. J. B} \textbf{\bibinfo{volume}{31}},
  \bibinfo{pages}{29} (\bibinfo{year}{2003}).

\bibitem[{\citenamefont{Boettcher}(2010{\natexlab{a}})}]{Boettcher10a}
\bibinfo{author}{\bibfnamefont{S.}~\bibnamefont{Boettcher}},
  \bibinfo{journal}{Euro. Phys. J. B} \textbf{\bibinfo{volume}{74}},
  \bibinfo{pages}{363 } (\bibinfo{year}{2010}{\natexlab{a}}).

\bibitem[{\citenamefont{Campbell et~al.}(2004)\citenamefont{Campbell, Hartmann,
  and Katzgraber}}]{Campbell04}
\bibinfo{author}{\bibfnamefont{I.~A.} \bibnamefont{Campbell}},
  \bibinfo{author}{\bibfnamefont{A.~K.} \bibnamefont{Hartmann}},
  \bibnamefont{and} \bibinfo{author}{\bibfnamefont{H.~G.}
  \bibnamefont{Katzgraber}}, \bibinfo{journal}{Phys. Rev. B}
  \textbf{\bibinfo{volume}{70}}, \bibinfo{pages}{054429}
  (\bibinfo{year}{2004}).

\bibitem[{\citenamefont{Boettcher}(2005{\natexlab{b}})}]{EOSK}
\bibinfo{author}{\bibfnamefont{S.}~\bibnamefont{Boettcher}},
  \bibinfo{journal}{Eur.\ Phys.\ J.\ B} \textbf{\bibinfo{volume}{46}},
  \bibinfo{pages}{501} (\bibinfo{year}{2005}{\natexlab{b}}).

\bibitem[{\citenamefont{Aspelmeier et~al.}(2008)\citenamefont{Aspelmeier,
  Billoire, Marinari, and Moore}}]{Aspelmeier07}
\bibinfo{author}{\bibfnamefont{T.}~\bibnamefont{Aspelmeier}},
  \bibinfo{author}{\bibfnamefont{A.}~\bibnamefont{Billoire}},
  \bibinfo{author}{\bibfnamefont{E.}~\bibnamefont{Marinari}}, \bibnamefont{and}
  \bibinfo{author}{\bibfnamefont{M.~A.} \bibnamefont{Moore}},
  \bibinfo{journal}{J. Phys. A: Math. Theor.}
  \textbf{\bibinfo{volume}{41}}, \bibinfo{pages}{324008}
  (\bibinfo{year}{2008}).

\bibitem[{\citenamefont{Boettcher}(2010{\natexlab{b}})}]{Boettcher10b}
\bibinfo{author}{\bibfnamefont{S.}~\bibnamefont{Boettcher}},
  \bibinfo{journal}{J. Stat. Mech} p. \bibinfo{pages}{P07002}
  (\bibinfo{year}{2010}{\natexlab{b}}).

\bibitem[{\citenamefont{Palassini and Young}(2000)}]{Palassini00}
\bibinfo{author}{\bibfnamefont{M.}~\bibnamefont{Palassini}} \bibnamefont{and}
  \bibinfo{author}{\bibfnamefont{A.~P.} \bibnamefont{Young}},
  \bibinfo{journal}{Phys. Rev. Lett.} \textbf{\bibinfo{volume}{85}},
  \bibinfo{pages}{3017} (\bibinfo{year}{2000}).

\bibitem[{\citenamefont{Bollobas}(1985)}]{Bollobas}
\bibinfo{author}{\bibfnamefont{B.}~\bibnamefont{Bollobas}},
  \emph{\bibinfo{title}{Random Graphs}} (\bibinfo{publisher}{Academic Press},
  \bibinfo{address}{London}, \bibinfo{year}{1985}).

\bibitem[{\citenamefont{Hartmann and Rieger}(2004)}]{Dagstuhl04}
\bibinfo{editor}{\bibfnamefont{A.}~\bibnamefont{Hartmann}} \bibnamefont{and}
  \bibinfo{editor}{\bibfnamefont{H.}~\bibnamefont{Rieger}}, eds.,
  \emph{\bibinfo{title}{New Optimization Algorithms in Physics}}
  (\bibinfo{publisher}{Wiley-VCH}, \bibinfo{address}{Berlin},
  \bibinfo{year}{2004}).

\bibitem[{\citenamefont{Boettcher et~al.}(2008)\citenamefont{Boettcher,
  Katzgraber, and Sherrington}}]{Boettcher07b}
\bibinfo{author}{\bibfnamefont{S.}~\bibnamefont{Boettcher}},
  \bibinfo{author}{\bibfnamefont{H.~G.} \bibnamefont{Katzgraber}},
  \bibnamefont{and}
  \bibinfo{author}{\bibfnamefont{D.}~\bibnamefont{Sherrington}},
  \bibinfo{journal}{J. Phys. A: Math. Theor.}
  \textbf{\bibinfo{volume}{41}}, \bibinfo{pages}{324007}
  (\bibinfo{year}{2008}).

\bibitem[{\citenamefont{Parisi et~al.}(1993)\citenamefont{Parisi, Ritort, and
  Slanina}}]{parisi:93}
\bibinfo{author}{\bibfnamefont{G.}~\bibnamefont{Parisi}},
  \bibinfo{author}{\bibfnamefont{F.}~\bibnamefont{Ritort}}, \bibnamefont{and}
  \bibinfo{author}{\bibfnamefont{F.}~\bibnamefont{Slanina}},
  \bibinfo{journal}{J. Phys. A} \textbf{\bibinfo{volume}{26}},
  \bibinfo{pages}{247} (\bibinfo{year}{1993}).

\bibitem[{\citenamefont{Bouchaud et~al.}(2003)\citenamefont{Bouchaud, Krzakala,
  and Martin}}]{Bouchaud03}
\bibinfo{author}{\bibfnamefont{J.-P.} \bibnamefont{Bouchaud}},
  \bibinfo{author}{\bibfnamefont{F.}~\bibnamefont{Krzakala}}, \bibnamefont{and}
  \bibinfo{author}{\bibfnamefont{O.~C.} \bibnamefont{Martin}},
  \bibinfo{journal}{Phys. Rev. B} \textbf{\bibinfo{volume}{68}},
  \bibinfo{pages}{224404} (\bibinfo{year}{2003}).

\bibitem[{\citenamefont{Bray and Moore}(1984)}]{Bray84}
\bibinfo{author}{\bibfnamefont{A.~J.} \bibnamefont{Bray}} \bibnamefont{and}
  \bibinfo{author}{\bibfnamefont{M.~A.} \bibnamefont{Moore}},
  \bibinfo{journal}{J.~Phys.~C: Solid State Phys.}
  \textbf{\bibinfo{volume}{17}}, \bibinfo{pages}{L463} (\bibinfo{year}{1984}).

\bibitem[{\citenamefont{Boettcher and Percus}(2001)}]{Boettcher01a}
\bibinfo{author}{\bibfnamefont{S.}~\bibnamefont{Boettcher}} \bibnamefont{and}
  \bibinfo{author}{\bibfnamefont{A.~G.} \bibnamefont{Percus}},
  \bibinfo{journal}{Phys. Rev. Lett.} \textbf{\bibinfo{volume}{86}},
  \bibinfo{pages}{5211} (\bibinfo{year}{2001}).

\bibitem[{\citenamefont{Pal}(1996)}]{Pal96a}
\bibinfo{author}{\bibfnamefont{K.~F.} \bibnamefont{Pal}},
  \bibinfo{journal}{Physica A} \textbf{\bibinfo{volume}{223}},
  \bibinfo{pages}{283} (\bibinfo{year}{1996}).

\bibitem[{\citenamefont{Hartmann}(1997)}]{Hartmann97}
\bibinfo{author}{\bibfnamefont{A.~K.} \bibnamefont{Hartmann}},
  \bibinfo{journal}{Europhys. Lett.} \textbf{\bibinfo{volume}{40}},
  \bibinfo{pages}{429} (\bibinfo{year}{1997}).

\bibitem[{\citenamefont{Banavar et~al.}(1987)\citenamefont{Banavar, Bray, and
  Feng}}]{Banavar87}
\bibinfo{author}{\bibfnamefont{J.~R.} \bibnamefont{Banavar}},
  \bibinfo{author}{\bibfnamefont{A.~J.} \bibnamefont{Bray}}, \bibnamefont{and}
  \bibinfo{author}{\bibfnamefont{S.}~\bibnamefont{Feng}},
  \bibinfo{journal}{Phys. Rev. Lett.} \textbf{\bibinfo{volume}{58}},
  \bibinfo{pages}{1463} (\bibinfo{year}{1987}).

\bibitem[{\citenamefont{Hartmann and Young}(2001)}]{Hartmann01}
\bibinfo{author}{\bibfnamefont{A.~K.} \bibnamefont{Hartmann}} \bibnamefont{and}
  \bibinfo{author}{\bibfnamefont{A.~P.} \bibnamefont{Young}},
  \bibinfo{journal}{Phys. Rev. B} \textbf{\bibinfo{volume}{64}},
  \bibinfo{pages}{180404(R)} (\bibinfo{year}{2001}).

\bibitem[{\citenamefont{Wehr and Aizenman}(1990)}]{Wehr90}
\bibinfo{author}{\bibfnamefont{J.}~\bibnamefont{Wehr}} \bibnamefont{and}
  \bibinfo{author}{\bibfnamefont{M.}~\bibnamefont{Aizenman}},
  \bibinfo{journal}{J. Stat. Phys.} \textbf{\bibinfo{volume}{60}},
  \bibinfo{pages}{287} (\bibinfo{year}{1990}).

\bibitem[{\citenamefont{Parisi and Rizzo}(2008)}]{Parisi08}
\bibinfo{author}{\bibfnamefont{G.}~\bibnamefont{Parisi}} \bibnamefont{and}
  \bibinfo{author}{\bibfnamefont{T.}~\bibnamefont{Rizzo}},
  \bibinfo{journal}{Phys. Rev. Lett.} \textbf{\bibinfo{volume}{101}},
  \bibinfo{pages}{117205} (\bibinfo{year}{2008}).

\bibitem[{\citenamefont{Aspelmeier et~al.}(2003)\citenamefont{Aspelmeier,
  Moore, and Young}}]{Aspelmeier03}
\bibinfo{author}{\bibfnamefont{T.}~\bibnamefont{Aspelmeier}},
  \bibinfo{author}{\bibfnamefont{M.~A.} \bibnamefont{Moore}}, \bibnamefont{and}
  \bibinfo{author}{\bibfnamefont{A.~P.} \bibnamefont{Young}},
  \bibinfo{journal}{Phys. Rev. Lett.} \textbf{\bibinfo{volume}{90}},
  \bibinfo{pages}{127202} (\bibinfo{year}{2003}).

\end{thebibliography}

\end{document}